\documentclass[preprint,aps,pra,superscriptaddress,showpacs]{revtex4}
\usepackage{epsfig}

\begin{document}

\title{Tests of Local Position Invariance using Continuously Running Atomic Clocks}

\author{Steven~Peil}
\email{steven.peil@usno.navy.mil}
\author{Scott~Crane}
\altaffiliation{current address: Naval Research Laboratory, Washington, DC}
\author{James~L.~Hanssen}
\author{Thomas~B.~Swanson}
\author{Christopher~R.~Ekstrom}
\affiliation{United States Naval Observatory, Washington, DC 20392}

\date{\today}

\begin{abstract}

Tests of local position invariance (LPI) made by comparing the relative redshift of
atomic clocks based on different atoms have been carried out for a variety of pairs of
atomic species.  In most cases, several absolute frequency measurements per year are
used to look for an annual signal, resulting in tests that can span on order of a
decade.  By using the output of continuously running clocks, we carry out LPI tests
with comparable or higher precision after less than 1.5 years. These include new
measurements of the difference in redshift anomalies $\beta$ for hyperfine transitions
in $^{87}{\rm Rb}$ and $^{133}{\rm Cs}$ and in $^{1}{\rm H}$ and $^{133}{\rm Cs}$ and
a measurement comparing $^{87}{\rm Rb}$ and $^{1}{\rm H}$, resulting in a stringent
limit on LPI, $\beta_{\rm Rb} - \beta_{\rm H}=\left(-2.7 \pm 4.9\right) \times
10^{-7}$. The method of making these measurements for continuous clocks is discussed.

\end{abstract}

\maketitle


\section{Introduction}

Metric theories of gravity are based on the universal coupling of gravity to matter
and energy, formalized in the Einstein equivalence principle (EEP). This principle
states that gravitational acceleration is independent of composition (the weak
equivalence principle) and that nongravitational measurements should be independent of
the velocity of the freely-falling reference frame (local Lorentz invariance) and of
the location in spacetime (local position invariance; LPI) where they are carried out.
Most efforts to unify general relativity, the widely accepted metric theory of
gravity, with quantum field theory involve a breakdown of EEP~\cite{violation,
dilaton}.

According to conventional parametrizations, the principle of LPI is the least-well
tested tenet of EEP~\cite{lrr}. Furthermore, tests of LPI can be used to constrain the
coupling of certain fundamental constants to gravitational potential and more
generally to test for spatial dependence of the constants. A spatial dependence is
allowed under a particular scenario in string theory~\cite{dilaton}, and there have
been published claims of a measured spatial variation of the fine structure constant
from astronomical observations~\cite{alpha_astro}.

The traditional redshift for a clock with frequency $\nu$ depends only on the change
in gravitational potential $\Delta U$, $\Delta \nu /\nu = \Delta U/c^2$, and should be
independent of the clock composition, for example. Violation of LPI may manifest
itself in an anomalous gravitational redshift of an atomic clock, commonly
parameterized with a clock-dependent term $\beta$:
\begin{equation}
\Delta \nu /\nu=(1+\beta)\Delta U/c^2.
\end{equation}
Measurements of gravitational redshift vs $\Delta U$ provide a test of the redshift
formula and of LPI~\cite{GPA}. Alternatively, the relative redshift of two clocks of
different composition can be used; if the two clocks have LPI-violating parameters
$\beta_1$ and $\beta_2$, the relative frequencies should vary with gravitational
potential as
\begin{equation}
\left( \Delta \nu /\nu\right)_{1,2}=(\beta_1-\beta_2)\Delta U/c^2. \label{e.betas}
\end{equation}
This differential measurement can be made with a higher precision than a direct
measurement of redshift vs $\Delta U$ at the possible expense of a reduction in the
size of the effect if $\beta_1$ and $\beta_2$ are close in value. The gain in
precision reported here compared to direct redshift measurements is about 200,
indicating that the two redshift anomalies must be the same to $\sim 5\times 10^{-3}$
for the best LPI test by absolute measurement of the gravitational redshift to be
comparable~\cite{GPA}. The possible range of values for $\beta_1 - \beta_2$ is one
reason why tests with many different pairs of clocks is valuable.

A convenient gravitational potential to use for an LPI test is the solar potential
experienced on earth; due to the earth's elliptical orbit, this varies annually as
\begin{equation}
\Delta U_s/c^2= A \sin(\omega t + \phi_0), \label{e.gravpot}
\end{equation}
where $A=1.65\times 10^{-10}$, $\omega = 0.0172$~rad/day, and $\phi_0$ is the phase
such that $\Delta U_s$ is a minimum at aphelion, such as modified Julian day 55746
(July 4, 2011).  (The additional variation in potential due to the earth's rotation
about its axis is $\sim 10^{-3} \Delta U_s$ for our latitude and is ignored here.) The
difference in LPI-violating parameters for two atomic clocks of different composition
can be measured by determining the size of the annual oscillation with the appropriate
phase in the clocks' relative frequency.

\begin{figure}
\includegraphics[width=0.5\textwidth]{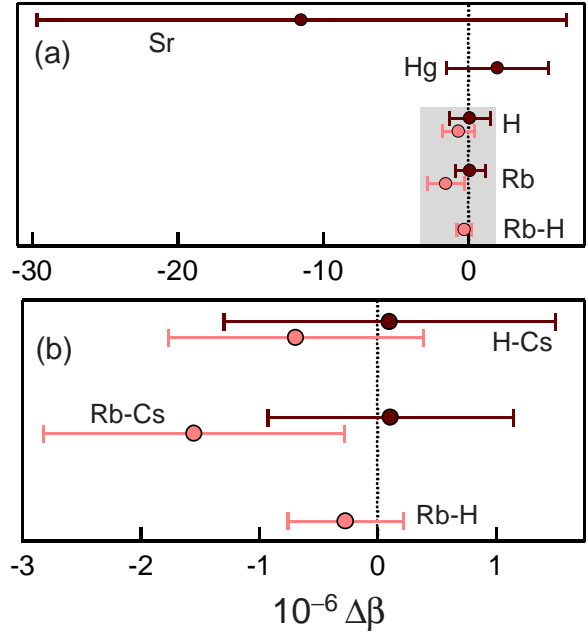}
\caption{(Color online.) LPI tests for various pairs of atomic species in terms of the
difference in anomalous redshift parameters, $\Delta \beta=\beta_1 - \beta_2$. (a)
Dark data points are previous measurements: (i) neutral strontium optical transition
against cesium standard, carried out over 2.5~years~\cite{strontium}\cite{stront2};
(ii) mercury ion optical transition against cesium standard, carried out over
5~years~\cite{mercury}; (iii) hydrogen hyperfine transition against cesium standard,
carried out over 7~years~\cite{ashby}; and (iv) rubidium-87 hyperfine transition
against cesium standard, carried out over 14~years~\cite{paris}.  Light data points
are results presented in this work. (b) A more detailed look at the shaded region in
(a).} \label{f.past_measurements}
\end{figure}

Over the past decade measurements of this sort have been made using a variety of pairs
of atomic species~\cite{dysprosium, mercury, ashby, strontium, paris},
summarized in Fig.~\ref{f.past_measurements}. In almost all previous measurements
absolute frequencies of a certain type of atomic clock are compared to a cesium
standard over many years, with typically several (fractional) frequency comparisons at
the $\sim 10^{-15}$ level per year. We implement a different approach, where each of
the atomic clocks used is in continuous operation, providing a much higher rate of
frequency comparisons and resulting in a more efficient LPI test. We measure the
output frequencies of clocks based on hyperfine transitions in $^{87}{\rm Rb}$,
$^{133}{\rm Cs}$, and $^{1}{\rm H}$ over 1.5 years.  Frequency comparisons between any
two of the atomic species can be made with a statistical uncertainty of $\sim
10^{-15}$ in a day or two, resulting in several hundred data points per year of
precision comparable to previous tests.

\section{Clocks, Method and Analysis}

Because we do not evaluate absolute frequencies, we have to be confident that our
clocks' outputs are stable or change in knowable, predictable ways.  This is
accomplished by maintaining a continuous record of a clock's frequency against a
variety of other clocks, enabling us to determine if discontinuous changes in
frequency or drift rate occur.

The US~Naval Observatory (USNO) has a large ensemble of atomic clocks used for precise
time~\cite{matsakis}, including commercial cesium-beam clocks, hydrogen masers, and
recently added rubidium-fountain clocks, which have been in operation for 1.5
years~\cite{peil}. All clocks are measured regularly against a master reference, which
enables a comparison between any two clocks. The collection of each type of clock can
be used to generate one or more stationary outputs representing that ensemble.

The rubidium frequency record used for LPI tests is an average of the frequencies of
the two highest performing fountains over the past 1.5 years. Comparisons among
rubidium fountains demonstrate that these two clocks exhibit extremely stable
frequencies with white-frequency noise levels below $2\times 10^{-13}$. The fountains
show no indication of frequency drift at the level of $3\times 10^{-18}$/day, as
determined by using cesium-fountain primary standards that contribute to International
Atomic Time as a reference.  As a baseline, we look for an LPI-violating signal
in the frequency comparison of the two rubidium fountains.  The result 
is 0 well within the error bar, indicating no bias in measurements using these clocks.

About 70 commercial cesium clocks are intercompared to create a single output used for
our LPI analysis~\cite{primaries}.  The output of this cesium ensemble is
characterized by a white-frequency noise level of $1\times 10^{-12}$ and shows a drift
in fractional frequency of $2\times 10^{-17}$/day over the period in question. There
is no long frequency record for an individual cesium clock without adjustments based
on the performance of the ensemble, so there is no way to measure an LPI-violating
baseline for the individual clocks.

Hydrogen masers generally exhibit complicated behavior over long averaging times
(months). They are known to be sensitive to environmental factors, which can cause
frequency discontinuities in addition to less violent perturbations, and they exhibit
significant frequency drifts, which may not be linear and which can change in time. Of
the $\sim25$ hydrogen masers available to us, a subset is ruled out for these LPI
tests because they exhibit discontinuous changes in frequency or drift rate during the
interval of time that we are interested in. Further assessing which masers to include
requires careful analysis of the frequency records of different masers compared
against each other as well as against the rubidium and cesium clocks used in
the analysis, as discussed further below.  
The maser short-term frequency stability is better than that of either of the other
types of clock used for these LPI tests.  The white-frequency noise level of a
maser-fountain comparison is typically of the order of $5\times 10^{-13}$, limited by
noise in the frequency-counter measurement system.  Some masers are also measured with
a dual-mixer measurement system, in which case a maser-fountain comparison exhibits a
white-frequency noise level limited by fountain performance.

In order to measure $\beta_{\rm 1} - \beta_{\rm 2}$ for a pair of atomic species we
need to examine the relative frequency record for an annual oscillation with phase
$\phi_0$ in the presence of a drift. We limit our analysis to clocks that show a
frequency drift that is linear, which, for a $\sim1.5$-year interval, has little
impact on the determination of the sinusoid amplitude.  Applying a least-squares fit
using a function that is a combination of drift and oscillation gives results
identical to fitting the frequency record to a line and then fitting those residuals
to a sinusoid. We use the latter procedure for presenting data in the figures below.
All relative
frequency measurements participating in the fit derive from averaging intervals 
resulting in white frequency noise, consistent with application of least-squares curve
fitting.

\subsection{Rubidium vs Cesium}

\begin{figure}
\includegraphics[width=0.5\textwidth]{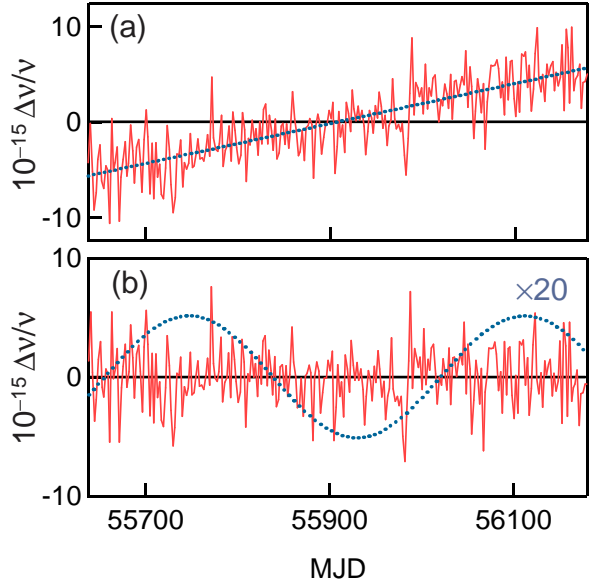}
\caption{(Color online.) (a) Frequency difference of rubidium-fountain average and
cesium ensemble over 1.5 years.  Fit to a linear drift is shown.  (b) Residuals after
removing drift, along with the fit to an LPI-violating signal, amplified $\times20$.
MJD stands for modified Julian day.} \label{f.rb-cs}
\end{figure}

Application of this procedure to a rubidium-cesium LPI test is straightforward.
Figure~\ref{f.rb-cs}(a) shows the fractional frequency difference between the cesium
ensemble and the average of rubidium fountains over 1.5 years. Raw data are recorded
at hourly intervals; the plots in the figure show the frequency averaged over 2 days.
The graph includes a fit indicating a relative drift of $2\times 10^{-17}$/day. In
Fig.~\ref{f.rb-cs}(b), the residuals after removing the drift are fit to an
LPI-violating signal, shown amplified in the graph. The amplitude for this fit is
$\Delta \nu_{\rm Rb}/\nu_{\rm Rb}-\Delta \nu_{\rm Cs}/\nu_{\rm Cs}= \left( 2.6 \pm 2.1
\right) \times 10^{-16}$ (all uncertainties are 1 standard deviation). This translates
to a difference in redshift anomalies of
\begin{equation}
\beta_{\rm Rb} - \beta_{\rm Cs} = \left( -1.6 \pm 1.3 \right) \times 10^{-6}.
\end{equation}

\subsection{Hydrogen}

From the subset of masers that exhibit no apparent discontinuities in frequency or
drift over all or a large portion of the past 1.5 years, we need to establish which
masers have frequency records that are truly linear, without higher order terms. Some
masers obviously exhibit frequencies that change in time in a nonlinear way and can be
removed from consideration.  To rule out masers with a more subtle nonlinear behavior,
we rely on the quality of the least-squares fit to the relative frequency record of a
maser and the rubidium average~\cite{rubidium}. If the data do not fit well, as
indicated by the reduced $\chi^2$ and a visual inspection of the residuals, the maser
is not included in the analysis.  We attribute a poor fit to unmodeled behavior in the
maser; this procedure leaves us with four masers in the LPI analysis.

It is possible that these four masers still display unmodeled behavior, which mimics
an LPI-violating signal.  Such unmodeled behavior would not be surprising since the
phase of the LPI-violating signal is close to that of the seasonal cycle. We asses
this by looking at all six possible pairs of the four masers for an LPI-violating
signal. These measurements enable a $\beta$ to be determined for each maser that
serves as a baseline for remaining unmodeled behavior.  Each $\beta$ value serves as a
bias and is subtracted from the measured LPI-violating signal for each maser-rubidium
or maser-cesium comparison. Once this bias is included in a clock comparison using a
given maser, the final uncertainty for that comparison is the combined statistical
uncertainty from the LPI fit and the uncertainty associated with the bias.

\begin{figure}
\includegraphics[width=0.5\textwidth]{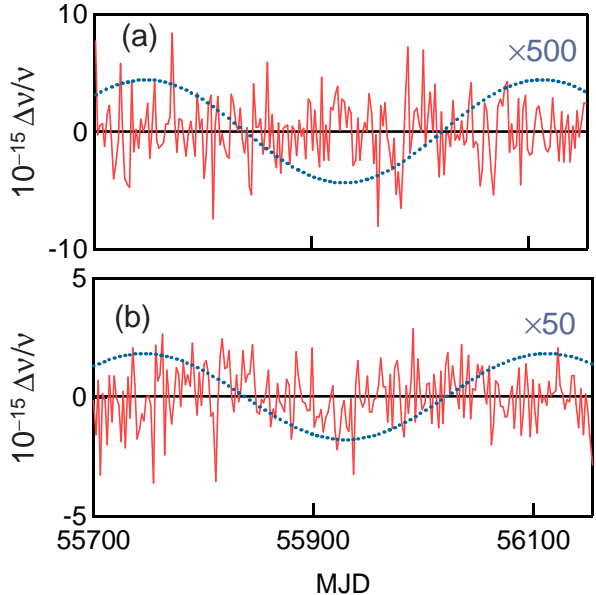}
\caption{(Color online.) Frequency comparison for one hydrogen maser and (a) the
cesium ensemble and (b) the rubidium-fountain average. Each point is a 2-day average,
and linear drifts have been removed.  Fits to LPI-violating signals are amplified by
the factors shown.} \label{f.hydrogen}
\end{figure}

In order to account for the different uncertainties associated with each maser, the
hydrogen results are determined by measuring LPI-violating amplitudes for each of the
four masers against the cesium and rubidium clocks and combining them in a weighted
average~\cite{uncertainty}.  The four measurements against the cesium ensemble
give
\begin{equation}
\beta_{\rm H} - \beta_{\rm Cs}=\left( -0.7 \pm 1.1 \right) \times 10^{-6}.
\end{equation}
The average of the rubidium fountains measured against the four masers yields
\begin{equation}
\beta_{\rm Rb} - \beta_{\rm H}=\left( -2.7 \pm 4.9 \right) \times 10^{-7}.
\end{equation}
Figure~\ref{f.hydrogen} shows plots of the frequency of one of the masers compared
against [Fig.~\ref{f.hydrogen}(a)] the cesium ensemble and [Fig.~\ref{f.hydrogen}(b)]
the rubidium average.

\section{Discussion and Summary}

Measurements of $\beta_{1} - \beta_{2}$ for different pairs of atomic species can be
used to constrain the coupling to gravitational potential of certain dimensionless
constants: the fine structure constant $\alpha$, the electron-to-proton mass ratio
$m_e/m_p$, and the ratio of the light quark mass to the quantum chromodynamics length
scale, $m_q/\Lambda_{\rm QCD}$, where $m_q$ is the average of the up and down quark
masses~\cite{flambaum}.  For the case of massive bodies the size of the sun, the
coupling of these constants to gravity is expected to be proportional to the coupling
to a hypothetical scalar field that is a component of many cosmological models and
that could affect their values~\cite{dilaton, flambaum}.

\begin{figure}
\includegraphics[width=0.5\textwidth]{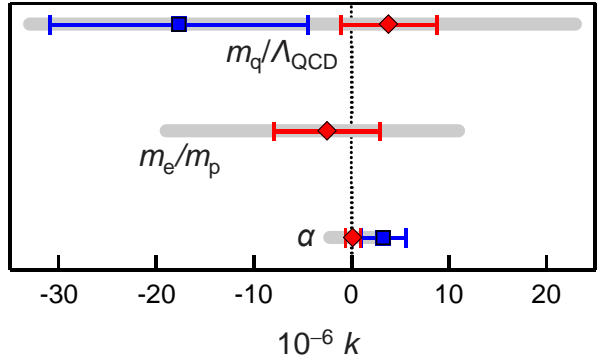}
\caption{(Color online.) Constraints on coupling of $\alpha$, $m_e/m_p$ and
$m_q/\Lambda_{\rm QCD}$ to gravitational potential characterized by the dimensionless
parameters $k_\alpha$, $k_{m_e/m_p}$, and $k_{m_q/\Lambda_{\rm QCD}}$. Gray bands show
constraints imposed by previous measurements, compiled from Refs.~\cite{mercury,
ashby, strontium, paris} and presented in~\cite{paris}. Diamonds show the tighter
constraints that result when our measurements are included in the analysis.  Squares
represent the values for $k_\alpha$ and $k_{m_q/\Lambda_{\rm QCD}}$ that are derived
using only the results presented here.} \label{f.constants}
\end{figure}

The coupling of these fundamental constants to gravitational potential can be
characterized by dimensionless coupling parameters $k_{\epsilon}$~\cite{flambaum}:
\begin{equation}
\frac{\delta \epsilon}{\epsilon} = k_{\epsilon} \left( \frac{\Delta U}{c^2} \right),
\end{equation}
where $\epsilon$ stands for the three constants $\alpha$, $m_e/m_p$ and
$m_q/\Lambda_{\rm QCD}$. Using calculated sensitivities of different atomic transition
frequencies to variations in these constants, differences in redshift anomalies for a
pair of atomic species can be related to linear combinations of the coupling
parameters $k_{\epsilon}$~\cite{paris, flambaum2, flambaum3}. The three LPI
measurements presented here can be used to put limits on the coupling of $\alpha$ and
$m_q/\Lambda_{\rm QCD}$ using only clocks from one institution, $k_\alpha=\left(
3.3\pm 2.3 \right) \times 10^{-6}$ and $k_{m_q/\Lambda_{\rm QCD}}=\left( -1.8 \pm 1.3
\right) \times 10^{-5}$. Measurements between optical and microwave clocks are needed
to constrain the coupling for $m_e/m_p$. By using our measurements along with those
from references~\cite{mercury, ashby, strontium, paris} we get limits on all three
couplings:
\begin{eqnarray}
k_{\alpha} &=& \left( 1.7\pm 7.5 \right) \times 10^{-7} \nonumber \\
k_{m_e/m_p} &=& \left( -2.5\pm 5.4 \right) \times 10^{-6} \nonumber \\
k_{m_q/\Lambda_{\rm QCD}} &=& \left( 3.8\pm 4.9 \right) \times 10^{-6}.
\end{eqnarray}
These results are plotted in Fig.~\ref{f.constants}.

The tests involving cesium are limited by the white-frequency noise of the cesium
ensemble, so the precision of these measurements should improve with longer fitting
times.  The limit on the precision of the rubidium-hydrogen test is a combination of
white-frequency noise from the measurement system and uncertainty in the LPI-mimicking
amplitude for each maser. While these uncertainties should also improve with longer
fitting times, the unpredictable nature of masers makes this less certain. Even now,
with an uncertainty on the rubidium-hydrogen measurement in the mid $10^{-7}$s, this
very precise LPI test could remain competitive for the near-future. While relative
frequency measurements of two optical clocks should yield much more stringent limits,
many optical clock frequencies are measured against the cesium standard, limiting the
precision to a level governed by a microwave transition. Furthermore, our continuous
clock measurements are competitive with LPI tests that could be implemented in
near-term space missions. Clock experiments intended for the International Space
Station~\cite{ACES} could improve upon the precision of absolute redshift measurements
but not differential measurements; the driving term due to the solar potential is the
same as on earth, and there is no driving term from the earth's potential due to the
station's circular orbit.  In terms of space clocks, improvements to differential
redshift tests of LPI could result from proposed missions that would put clocks in a
highly eccentric earth orbit~\cite{EGE} or a solar system escape
trajectory~\cite{SAGAS}.

In summary, we have made precise LPI tests with three different pairs of atomic
species using continuously running atomic clocks and have used these to tighten the
constraints on coupling of fundamental constants to gravity.

\section{Acknowledgements}
We are indebted to Paul Koppang for providing clock data and an understanding of the
cesium ensemble, and we benefited from discussions with Demetrios Matsakis. Atomic
fountain development at USNO has been funded by ONR and SPAWAR.

\bibliographystyle{prsty}

\begin{thebibliography}{10}
\bibitem{stront2}
The strontium uncertainty in the plot is a factor of 2 larger than published
in~\cite{strontium}.  See also \cite{paris}.
\bibitem{primaries}
Using the commercial cesium clocks at USNO is more straightforward than using the
cesium fountains that contribute to International Atomic Time, although measurements
against the world's primary standards may be considered in future work.
\bibitem{rubidium}
We rely on the comparison against a fountain because it is lower noise than the
comparison against the cesium average.
\bibitem{uncertainty}
For rubidium, individual maser measurements are uncorrelated. For cesium, correlation
between individual measurements gets taken into account.
\end{thebibliography}

\end{document}